\documentclass[twocolumn]{aastex631}
\usepackage{amsmath}
\usepackage{booktabs}
\usepackage{multirow}

\begin{document}

\title{Simulations of coronal mass ejections on a young solar-type star and their detectability through coronal spectral observations}

\author[0000-0002-7421-4701]{Yu Xu}
\affiliation{School of Earth and Space Sciences, Peking University, Beijing 100781, China; xuyu@stu.pku.edu.cn, huitian@pku.edu.cn}
\affiliation{Leibniz Institute for Astrophysics Potsdam, An der Sternwarte 16, D-14482 Potsdam, Germany}

\author[0000-0002-1369-1758]{Hui Tian}
\affiliation{School of Earth and Space Sciences, Peking University, Beijing 100781, China; xuyu@stu.pku.edu.cn, huitian@pku.edu.cn}
\affiliation{State Key Laboratory of Solar Activity and Space Weather, National Space Science Center, Chinese Academy of Sciences, Beijing 100190, China}

\author[0000-0001-5052-3473]{Juli\'{a}n D. Alvarado-G\'{o}mez}
\affiliation{Leibniz Institute for Astrophysics Potsdam, An der Sternwarte 16, D-14482 Potsdam, Germany}

\author[0000-0002-0210-2276]{Jeremy J. Drake}
\affiliation{Lockheed Martin, 3251 Hanover St, Palo Alto, CA 94304, USA}

\author[0000-0002-2671-8796]{Gustavo Guerrero}
\affiliation{Universidade Federal de Minas Gerais Av. Antonio Carlos, 6627, Belo Horizonte, MG 31270-901, Brazil}

\begin{abstract}
There is a growing interest in searching for coronal mass ejections (CMEs) in other stellar systems because they are thought to be one of the important factors {shaping planetary atmospheres}. We investigated the possible spectral signatures related to stellar CMEs using magnetohydrodynamic simulations and spectral synthesis techniques. Blue wing enhancements of the synthetic coronal line profiles caused by the line-of-sight motion of plasma were observed during the simulated CME events. We included instrumental conditions in the spectral synthesis and tested the detectability of the asymmetries under different instrumental broadening conditions. {
The results show that blue wing asymmetries are visible in some EUV lines with spectral resolutions higher than around 2000, and the line-of-sight velocities of CMEs obtained from asymmetry analysis techniques are comparable to the CME velocities derived from three-dimensional model outputs. However, when the spectral resolution drops below 2000, the asymmetries in the blue wings become barely visible, but blue shifts in the line centroids with velocities around -100 to -200 km/s are observed.} {We suggest a method of using MHD simulation to synthesize line profiles and analyse their asymmetries which} may help to guide future instrument design in terms of detecting stellar CMEs through Doppler shifts or asymmetries of coronal spectral lines.
\end{abstract}

\section{introduction}\label{sec:introduction}
Stellar coronal mass ejections (CMEs) have drawn increasing attention due to the growing realization of their potential effects on planetary habitability (e.g., \citealt{Airapetian2016,Alvarado-Gomez2022,Hazra2022}). Stellar CMEs propagate from the stellar surface to the surrounding interplanetary space, altering the magnetic configuration, density, temperature, and other physical properties by their passage, adding to and possibly dominating the so-called space weather conditions.

CMEs on the Sun are often observed by coronagraphs in white-light, for instance, by the Large Angle Spectroscopic Coronagraph (LASCO; \citealt{Brueckner1995}) onboard the Solar and Heliospheric Observatory (SOHO; \citealt{Domingo1995}). CMEs can also be observed at extreme-ultraviolet (EUV) wavelengths. For example, the Atmospheric Imaging Assembly (AIA; \citealt{Lemen2012}) on board the Solar Dynamics Observatory (SDO; \citealt{Pesnell2012}) and the Solar Upper Transition Region Imager (SUTRI; \citealt{Bai2023}) on board the Space Advanced Technology demonstration satellite (SATech-01) can capture the early phase of CME propagation. However, in the case of other stars spatially resolved observations are unavailable due to the large distances of stellar systems.  Current stellar observations only provide disk-integrated spectra or photometry. In order to understand CME signatures in stellar observations, recourse has been sought in potential CME indicators in full-Sun spectroscopic observations. Coronal dimmings related to the mass loss process during CMEs are a well-known phenomenon and have been characterized in detail (e.g., \citealt{Mason2014,Mason2016}). Obscuration dimmings due to absorption by dense filament material during filament eruptions have also been reported and shown to be detectable in full-disk He II 304~\AA\ emission \citep{Xu2024}. Doppler shifts caused by the line-of-sight (LOS) motions of plasma were found during solar CME events (e.g., \citealt{Xu2022,Lu2023,Otsu2024}). These signatures are interpreted as indicators of either CMEs or filament eruptions and have been sought in stellar observations. Candidate coronal dimmings on stars have been mainly found in X-ray and far-ultraviolet wavelengths (e.g., \citealt{Veronig2021,Loyd2022}). X-ray absorptions have been detected during stellar flare events (e.g., \citealt{Favata1999,Moschou2017,Wang2023}). Doppler shifts related to mass ejections or filament eruptions were observed in line profiles of spectral lines at various wavelengths (e.g., \citealt{Houdebine1990,Fuhrmeister2004,Leitzinger2011,Argiroffi2019,Vida2019,Lu2022,Namekata2022,Chen2022,Inoue2024a,Inoue2024b}).

Although great progress has been made in stellar CME observations, there are still lots of mysteries. For example, the dimming depths identified on other stars are much larger than those on our Sun, which requires more comprehensive explanation. The plasma motions during flare processes can also cause asymmetries in line profiles, which brings ambiguities to the interpretation of the Doppler shifts in stellar observations. CME masses and velocities have been inferred from stellar observations \citep[see, e.g.,][]{Moschou2017,Moschou2019} but with large uncertainties. 

Magnetohydrodynamic (MHD) simulations are another method to investigate stellar CMEs. Simulations of CMEs on solar-type stars and other late-type stars have been reported in previous studies, improving our understanding of stellar CME properties and their propagation processes (e.g., \citealt{Lynch2019}), their low detection rate (e.g., \citealt{Alvarado-Gomez2018,Alvarado-Gomez2019,Alvarado-Gomez2020}), and their interaction with the planetary atmosphere (e.g, \citealt{Cohen2022,Fraschetti2022,Hu2022}).

Here, we aim at investigating disk-integrated spectra during stellar CMEs through MHD simulations. We launched several CME cases on a solar-type star using a solar MHD model, and conducted spectral synthesis to investigate the evolution of the profiles of several spectral lines at EUV and FUV wavelengths. Spectral synthesis was based on the time-dependent outputs of the model physical parameters at each grid point in the three-dimensional (3D) simulation space. The basic parameters of the star (i.e., mass $M_\bigstar=1.16~\mathrm{M_\odot}$, stellar radius $R_\bigstar=1.23~\mathrm{R_\odot}$, rotation period $P_\bigstar=7.7~\mathrm{d}$) were inspired by a young solar-type star named $\iota$ Horologii \citep{Bruntt2010,Alvarado-Gomez2018b}. The inner boundary condition of the model is a magnetogram adopted from a stellar dynamo model (\citealt{Guerrero2019}, see Section \ref{sec:methods} for details). We analyzed the asymmetries in the synthetic spectra and provided constraints on the type of observations for their detection.

The paper is arranged in the following manner: Section \ref{sec:methods} briefly introduces the MHD model and describes the spectral synthesis techniques as well as the methods used to analyze the asymmetries in line profiles. Section~\ref{sec:results} presents the results, including the physical parameters of the simulated CMEs and the evolution of the synthetic spectra. Section \ref{sec:discussion} and \ref{sec:summary} present a discussion and a summary of the findings, respectively.

\section{Method}\label{sec:methods}
\subsection{The AWSOM and CME Models}

The Alfv\'{e}n Wave Solar Model (AWSOM; \citealt{Sokolov2013,Holst2014}) is part of the Space Weather Modeling Framework (SWMF; \citealt{Toth2012}) developed and maintained by the research team at the University of Michigan. {The model uses the BATS-R-US code \citep{Powell1999} to solve MHD equations in discrete 3D space}. It was initially designed for the Sun and has been validated by comparing the model outputs with remote sensing or in-situ observations (e.g., \citealt{Cohen2009,Downs2012,Jin2013,Manchester2014,Jin2017,Li2021}). The code has also been widely used in stellar simulations, including simulations of stellar wind conditions, transient events like CMEs, and star-planet interactions (e.g., \citealt{Alvarado-Gomez2016a,Alvarado-Gomez2016b,doNascimento2016,Garraffo2016,Dong2018,Folsom2020,Liu2022,Chebly2023,Xu2024b}).

\begin{figure*}
     \centering
\includegraphics[width=0.9\textwidth]{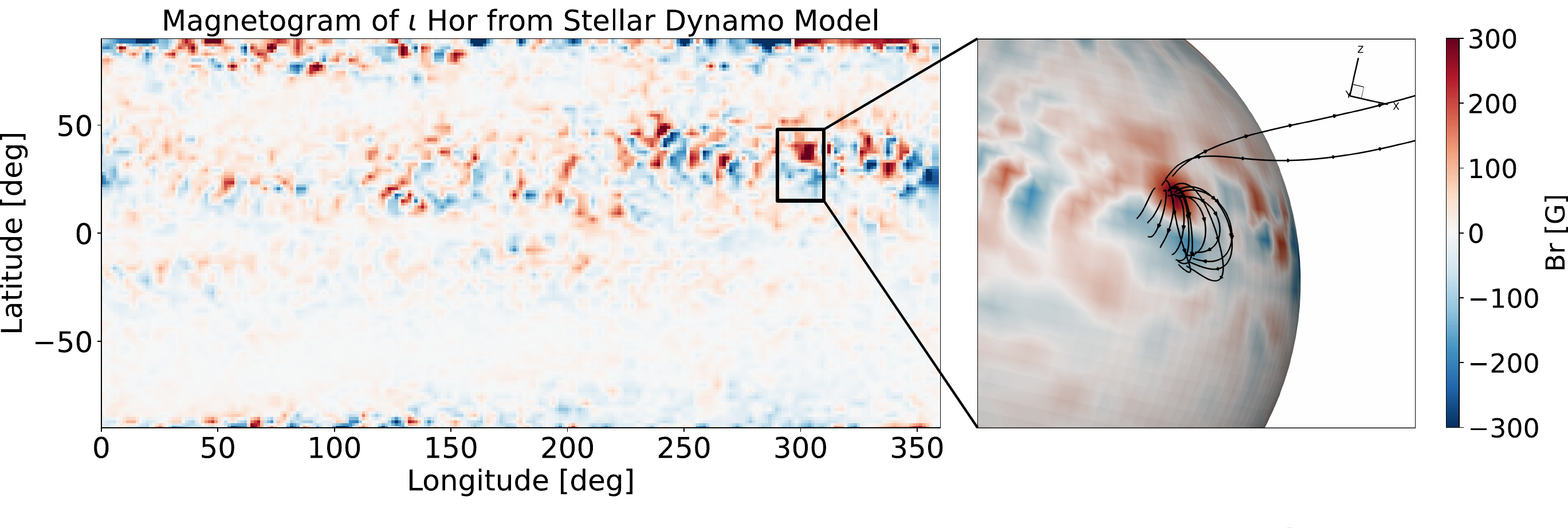}
     \caption{The magnetogram adopted from \cite{Guerrero2019} and used as the inner boundary condition of our simulations. The box outlines the bipolar region where the flux ropes were inserted to represent the CME. The region is projected on a sphere for visualization and some field lines are traced.}
    \label{fig:magnetogram}
\end{figure*}

The model inputs also include some parameters related to coronal heating. We adopted the following values for our simulations: the temperature and density at the top of the chromosphere are $5\times10^4\mathrm{K}$ and $2\times 10^{10}$~cm$^{-3}$, the Alfv\'{e}n wave correlation length $L_{\perp}\sqrt{B}=1.5\times 10^{4}~\mathrm{m\cdot\sqrt{T}}$, and {the Poynting flux scaled by the magnetic field} $S_A/B=3.48\times 10^{6}~\mathrm{J\cdot m^{-2}\cdot s^{-1}\cdot T^{-1}}$. The values of the first two parameters, the density, and temperature at the top of the chromosphere, have been widely used in previous solar and stellar simulations (e.g., \citealt{Alvarado-Gomez2018,Alvarado-Gomez2019}). 

The last two parameters related to coronal heating, namely the Alfv\'{e}n wave correlation length and the Poynting flux scaled by the magnetic field strength, were constrained by the emission measure (EM) from  observations. The EM represents the plasma temperature distribution of the stellar corona in terms of its capacity to drive radiative loss and is given by  $EM(T)=\int_V N_e^2 dV$, where $N_e$ is the electron density, and $dV$ is the unit volume in the volume $V$ with temperature $T$. The EM as a function of temperature can be retrieved from spectroscopic observations and we took the observed EM curve of $\iota$ Hor from \citet{SanzForcada2019} as a reference. We performed 15 test runs with different combinations of $L_{\perp}\sqrt{B}$, $S_A/B$, and grid resolution, and calculated the EM from each of the outputs. We then compared the 15 simulated EM curves with the observed one and adopted the parameter set providing the best EM match. {\cite{SanzForcada2019} suggested that the two peaks in the observed EM curve (i.e., log T/K = 6.3 and 6.8) can be interpreted as loops with their maximum temperature at those two temperatures. We tried to adjust parameters in the model in order to align with the observed EM curve, but we could not find a solution with sufficient EM at high temperatures (i.e., around log T/K $>$ 6.0) and without large excess of the EM at low temperatures (i.e., around log T/K $<$ 5.5).  We selected the solution whose EM curve aligns with the observed one around the temperatures of log T/K =6.3 (see Sec. \ref{sec:results} for more details).}

The computational domain starts from the stellar surface and extends to $30~\mathrm{R_\bigstar}$ from the stellar center. Simulations were conducted with a spherical coordinate system. The longitudinal and latitudinal resolutions were $\sim1.4^o$ within $1.7~\mathrm{R_\bigstar}$, which is the same as the angular resolution of the magnetogram. The angular resolution was reduced to $\sim2.8^o$ outside $1.7~\mathrm{R_\bigstar}$ for higher computational efficiency. The grid size in the radial direction was uneven, with the finest grid having a resolution of $\sim 0.001~\mathrm{R_\bigstar}$. The model ran until a steady state was achieved, and then flux ropes were inserted to trigger the CME eruption.

We used the Gibson-Low flux rope model \citep{Gibson1998} to initialize the eruption. Its morphological properties can be described by several parameters, including the radius $R$, the apex height $h$, and the stretch $a$. The flux rope was placed in an active region with strong bipolar fields which is outlined in Fig. \ref{fig:magnetogram}. The location and orientation of the flux rope are controlled by three parameters, namely the longitude, latitude at the flux rope center, and the orientation angle of the flux rope. We used the parameter set of: $R=0.2~\mathrm{R_\bigstar}$, $h=0.4~\mathrm{R_\bigstar}$, $a=1.0~\mathrm{R_\bigstar}$, longitude of $301.1^o$, latitude of $20.1^o$, and the orientation angle of $165.9^o$. The choice of the parameters of the flux rope was aimed at mitigating the disturbance to the system caused by the insertion, and trying to align the flux rope perpendicular to the polarity inversion line (PIL) of the active region. 

We created three CME cases and the only difference among them was the initial field strength at the center of their flux ropes. The central field strengths of the three flux ropes were $20~\mathrm{Gauss}$ (Case 1), $50~\mathrm{Gauss}$ (Case 2), and $70~\mathrm{Gauss}$ (Case 3), respectively. These initial field strengths are comparable with the average strength of the surrounding magnetic field. The flux ropes were ejected right after the insertion, wrapping mass as well as magnetic fields together with them and propagating towards the interplanetary space. 

The total simulated evolution time of each case is 45 minutes, which is a compromise between capturing the initial stage of the eruptions and our limited computational resources. The 3D outputs from the simulations have a time cadence of 1 minute.

\subsection{EUV Images and Spectral Synthesis}

The eruption was visualized through synthetic EUV images at each time step, giving a straightforward impression of the morphology of the stellar corona and the temperature distribution of plasma. The SWMF includes a tool to synthesize images based on an instrumental response function and the direction of line-of-sight (LOS) defined by the user. We adopted response functions from the existing instruments SDO/AIA and HINODE/X-ray Telescope (HINODE/XRT; \citealt{Golub2007}). The wavebands we chose include seven SDO/AIA bands (i.e. $304$~\AA, $171$~\AA, $193$~\AA, $211$~\AA, $335$~\AA, $94$~\AA, and $131$~\AA) and one HINODE/XRT band $Al-Mesh$. The wavebands we adopted cover a wide range of temperatures spanning from around $10^5~\mathrm{K}$ to around $10^7~\mathrm{K}$. Each selected AIA band covers a relatively broad wavelength range. In the effective-area functions of AIA channels (see Fig. 8 in \citealt{Boerner2012}), the bandpass width is typically several angstroms, corresponding to an offset velocity of at least $\sim3500$ km/s. {The wavelength coverage of the $Al\text{-}Mesh$ band extends from approximately 4 to 70 angstroms, based on an effective area of 0.01 $\mathrm{cm^2}$ \citep{Golub2007}.}

The intensity of each pixel on the image of a certain waveband $w$ can be written as,
\begin{equation}
I_w=\int C(N_e,T)N_e^2\,dl
\label{eq:aiaxrt}
\end{equation}
where $C(N_e,T)$ is the instrumental response function, and $T$ and $N_e$ are the temperature and electron density, respectively. The $dl$ in the integration represents the unit length along the LOS direction. In the simulation outputs where the space is discrete, the integration is converted to a series sum.

The intensity of an optically thin line can be calculated by an equation similar to Eq. \eqref{eq:aiaxrt}. Substituting the instrumental response function with the contribution function of the ion emitting the line $i$, we obtain the intensity of the line by 
\begin{equation}
I_i=\int G(T)N_e^2\,dV
\label{eq:i}
\end{equation}
where $G(T)$ is the contribution function of the transition as a function of temperature. We adopted contribution functions obtained from the CHIANTI atomic database, with a fixed density of $10^{9}~\mathrm{cm^{-3}}$ \citep{DelZanna2021}. The contribution function can also depend on electron density for some density-sensitive lines, but in general the dependency on the electron density is much weaker than that on the temperature. With $dV$ being the unit volume, the integration was done over the whole computational domain {excluding the grid elements obscured by the stellar body}.

In observations, spectral lines are broadened by the thermal motion of the ion, the macro motion of the plasma, and the instrumental response function. The line intensity as a function of wavelength, $\lambda$, can be written as
\begin{equation}
I(\lambda)=\int\frac{G(T)N_e^2}{\sqrt{2\pi}\sigma}e^{-\frac{(\lambda-\lambda_0)^2}{2\sigma^2}}\, dV
\end{equation}
where $\lambda_0$ is the central wavelength in the unit volume $dV$ after including the plasma motion with  velocity $\vec{v}$ along the LOS $\vec{n}$, given by $\lambda_0=(1-\frac{\vec{v}\cdot\vec{n}}{c})\lambda_c$, with $\lambda_c$ being the rest wavelength of the line adopted from the CHIANTI database 10.1 \citep{DelZanna2021}. We integrated the entire volume of the simulation domain. 
$\sigma$ represents the line broadening and can be expanded thus
\begin{equation}
\sigma^2=(\frac{\lambda_0}{c})^2\bigg[\big(\sqrt{\frac{k_BT}{m_i}}\big)^2+v_{nth}^2+\sigma_{instru}^2\bigg],
\end{equation}
where $c$ is the speed of light, $k_B$
is the Boltzmann constant, and $m_i$ is the mass of the emitting ion. The first term in the brackets is the thermal broadening due to the micro motion of the ion. The non-thermal broadening term, $v_{nth}$, represents the broadening effect caused by the macro motion of the plasma, such as wave or turbulence. We adopted a typical value of $v_{nth}=30~\mathrm{km/s}$ from solar observations (e.g., \citealt{Chae1998}). The third part, $\sigma_{instru}$, represents the instrumental broadening effect due to the finite spectral resolution $R$ in reality. The instrumental broadening is calculated as $\sigma_{instru}=\frac{c}{R}$, where $c$ is speed of light. $R=\frac{\lambda}{\Delta\lambda}$, is a variable in the synthesis. The line profiles were normalized to their corresponding peak values, namely
\begin{equation}
F(\lambda)=\frac{I(\lambda)}{I_{max}}
\label{eq:lp}
\end{equation}
{In the following text, the independent variable in the function $F(\lambda)$ is also converted to the offset velocity $v$ relative to a reference wavelength $\lambda_{ref}$ by $v=\frac{\lambda-\lambda_{ref}}{\lambda_{ref}}c$.}

We selected ten prominent/coronal lines for investigation. Their formation temperatures range from $10^{5.8}~\mathrm{K}$ to $10^{7.3}~\mathrm{K}$. The lines, together with their wavelengths and peak formation temperatures under thermal equilibrium, are listed in Table \ref{tab:lines}. We varied the spectral resolution $R$ to investigate the asymmetries of line profiles under different instrumental conditions: we investigated $R = 500,~1000,~2000,~3000,~4000, and ~5000$. The situation where $R\to\infty$ (i.e. $\sigma_{instru}=0$) was used as a reference. We adopted a sampling velocity (or wavelength) related to the spectral resolution. The Nyquist criterion requires that the sampling frequency should be at least two times the highest frequency of the original signal. Here, we adopted a  sampling frequency of $2\sqrt2 \sigma_{instru}$. That is to say, if the spectral resolution has $\sigma_{instru} = 30~\mathrm{km/s}$, the sampling velocity in the synthetic line profiles will be around $10.6~\mathrm{km/s}$.

\begin{table}
    \centering
    \caption{Lines used in this study}
    \begin{tabular}{cccc}
        \hline
        \hline
        \multirow{2}{*}{Ions} & Rest Wavelength & Formation Temperature  \\
        & [\AA] & [log $T$/K] \\
        \hline
        \hline
        Fe~\sc{ix} & 171.000 & 5.81 \\
        \hline
        \multirow{2}{*}{Fe~\sc{xii}} & 1242.000 &\multirow{2}{*}{6.18}\\
        &1349.360&\\
        \hline
        Fe~\sc{xiv} & 211.317 & 6.27 \\
        \hline
        Fe~\sc{xvi} & 335.022 & 6.43 \\
        \hline
        Fe~\sc{xviii} & 93.932 & 6.81 \\
        \hline
        Fe~\sc{xix} & 108.355 & 6.95 \\
        \hline
        \multirow{2}{*}{Fe~\sc{xxi}} & 128.750 & \multirow{2}{*}{7.06} \\
        & 1354.100 &\\
        \hline
        Fe~\sc{xxiv} & 192.300 & 7.25 \\
        \hline
        \hline
        
    \end{tabular}
    \label{tab:lines}
\end{table}

\subsection{Asymmetry Analysis of Line Profiles}
The analysis of asymmetries in line profiles can be done in different ways, for instance, multi-Gaussian fitting (e.g., \citealt{Tian2012b,Chen2022}), red-blue analysis (RB analysis hereafter, see e.g., \citealt{DePontieu2009,DePontieu2010,Tian2012}), and subtracting the primary component from the line profile to obtain the shifted components (single Gaussian residual hereafter; see e.g., \citealt{Yang2022,Yang2024}). We chose the RB and single Gaussian residual techniques to analyze the asymmetries in the synthetic line profiles in this study.

The RB profile shows the asymmetries in either red or blue wings. It is defined by the equation
\begin{equation}
\begin{split}
RB(v)=&\int_{v}^{v+\Delta v} F(v) dv - \int^{-v}_{-v-\Delta v} F(v) dv,\\ 
v=& \frac{\lambda-\lambda_m}{\lambda_m}c,
\end{split}
\label{eq:rb}
\end{equation}
where $F(v)$ is the line profile (Eq. \eqref{eq:lp}) and $\Delta v$ is the sampling velocity. $\lambda_m$ is the wavelength corresponding to the maximum value in the line profile. The velocity $v$ in Eq. \eqref{eq:rb} is the offset velocity relative to $\lambda_m$. {Assuming the line profile consists of one blue-shifted Gaussian superimposed on a primary Gaussian located near the rest wavelength, the local minimum of the $RB(v)$ is supposed to correspond to the velocity of the blue-shifted component (denoted as $v^m_{rb}$). It is worth noting that the independent variable $v$ in Eq. \eqref{eq:rb} is defined relative to $\lambda_m$ and $\lambda_m$ is not necessarily equal to {the rest wavelength of the line $\lambda_c$}. We converted $v^m_{rb}$ to be the offset velocity relative to the rest wavelength of the line (denoted as $v_{rb}$) by $v_{rb}=\frac{\lambda_m}{\lambda_c}v_{rb}^m+\frac{\lambda_m-\lambda_c}{\lambda_c}c$}.

The single Gaussian residual method can also reveal asymmetries in line profiles. We first fitted the primary peak of the line profile using a Gaussian function. Then we subtracted the fitted Gaussian from the line profile. The residual showed how much the line profile deviates from a single Gaussian shape. We adopted the velocity corresponding to the local maximum of the residual in the blue wing to be the velocity of the blue-shifted component. {This method has been used in the previous works (e.g., \citealt{Yang2022,Yang2024}), and has been proven to retrieve the LOS velocity of the moving plasma under certain conditions.}

{The ideal scenario for using the aforementioned methods to diagnose CME velocities is one in which the background emission remains perfectly symmetric, and any superimposed blue-shifted component is entirely due to CME material. {In this case, the minimum RB value and the maximum Gaussian residual should align with the main peak of the blue-shifted component, and the velocity associated with this peak is adopted as a representative value, reflecting the bulk motion of the CME material at a certain temperature.} However, in reality the asymmetries in lines also have contributions from the surrounding wind and/or the plasma motion inside the flare region, making the derived velocities from the aforementioned methods also have contributions from these material. Also, the asymmetries are not necessarily significant enough to show the secondary component(s) in the line wings. Instead, the line is broadened during the event and the line centroid is shifted. Under this condition, the amplitude of the asymmetries may fall below the noise level, resulting large uncertainties in the velocity derivation.}

\section{Results}\label{sec:results}
\subsection{Steady State}
\begin{figure*}
    \centering
    \includegraphics[width=1.0\textwidth]{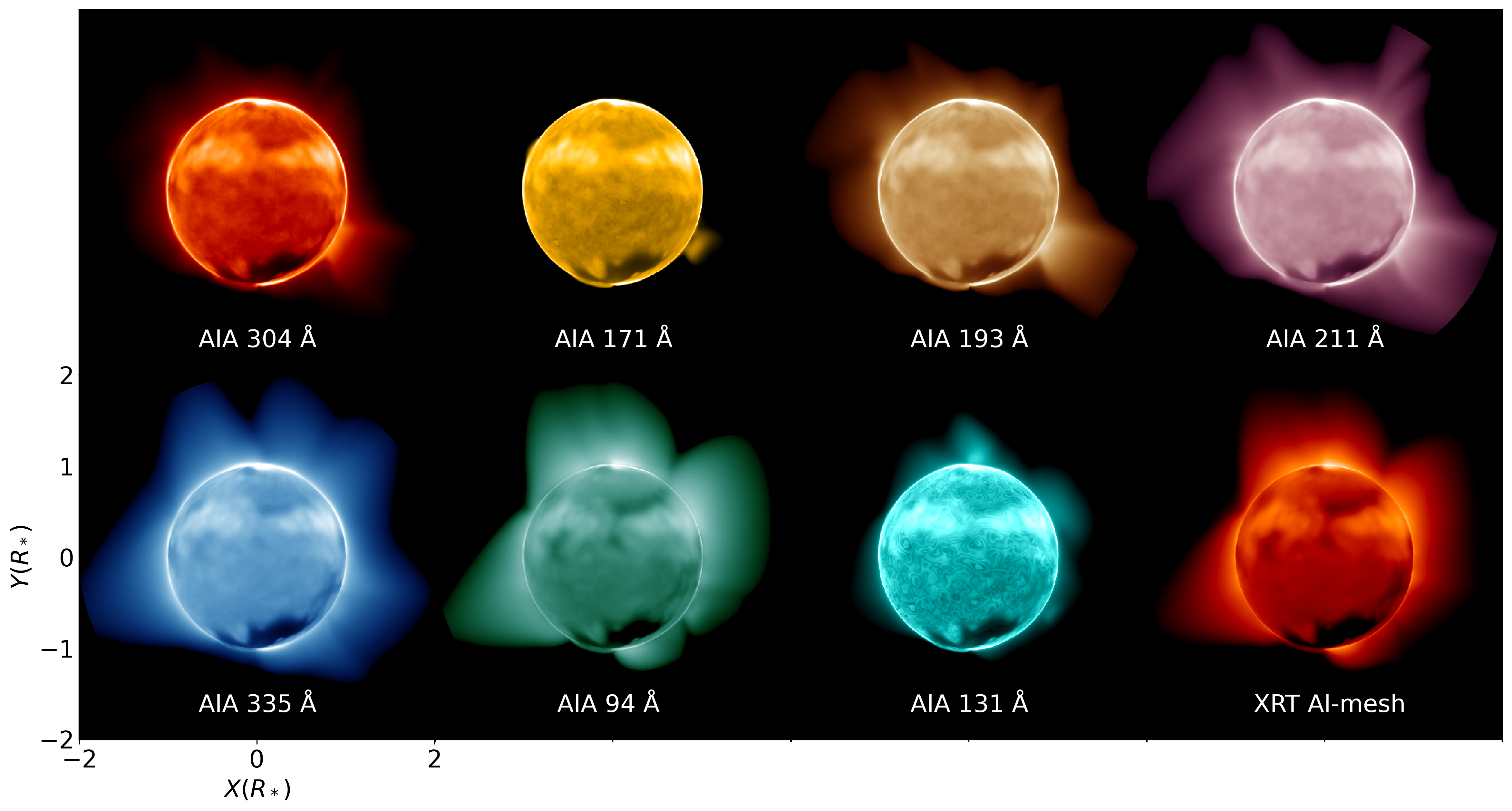}
    \caption{The synthesized AIA and XRT images of the steady-state corona labelled according to the bandpass name. {The images are in the arbitrary unit and logarithmic scale.}}
    \label{fig:euvss}
\end{figure*}

The steady-state solution produces a hotter corona than our sun, in broad accordance with observations \citep{SanzForcada2019}. Figure \ref{fig:euvss} displays the stellar corona in eight wavebands from a random viewing angle. A coronal hole in the southern polar region is visible in all wavebands. The extended corona can be seen in high-temperature channels. There is also a trace of an active region belt in the northern hemisphere, which corresponds to the regions with strong magnetic fields in the magnetogram. The contrast between the active regions and the quiescent regions is not as high as those we often see in solar observations, which suggests a lack of high temperature materials or an excess of low temperature materials in the simulated corona.

\begin{figure}
    \centering
    \includegraphics[width=0.49\textwidth]{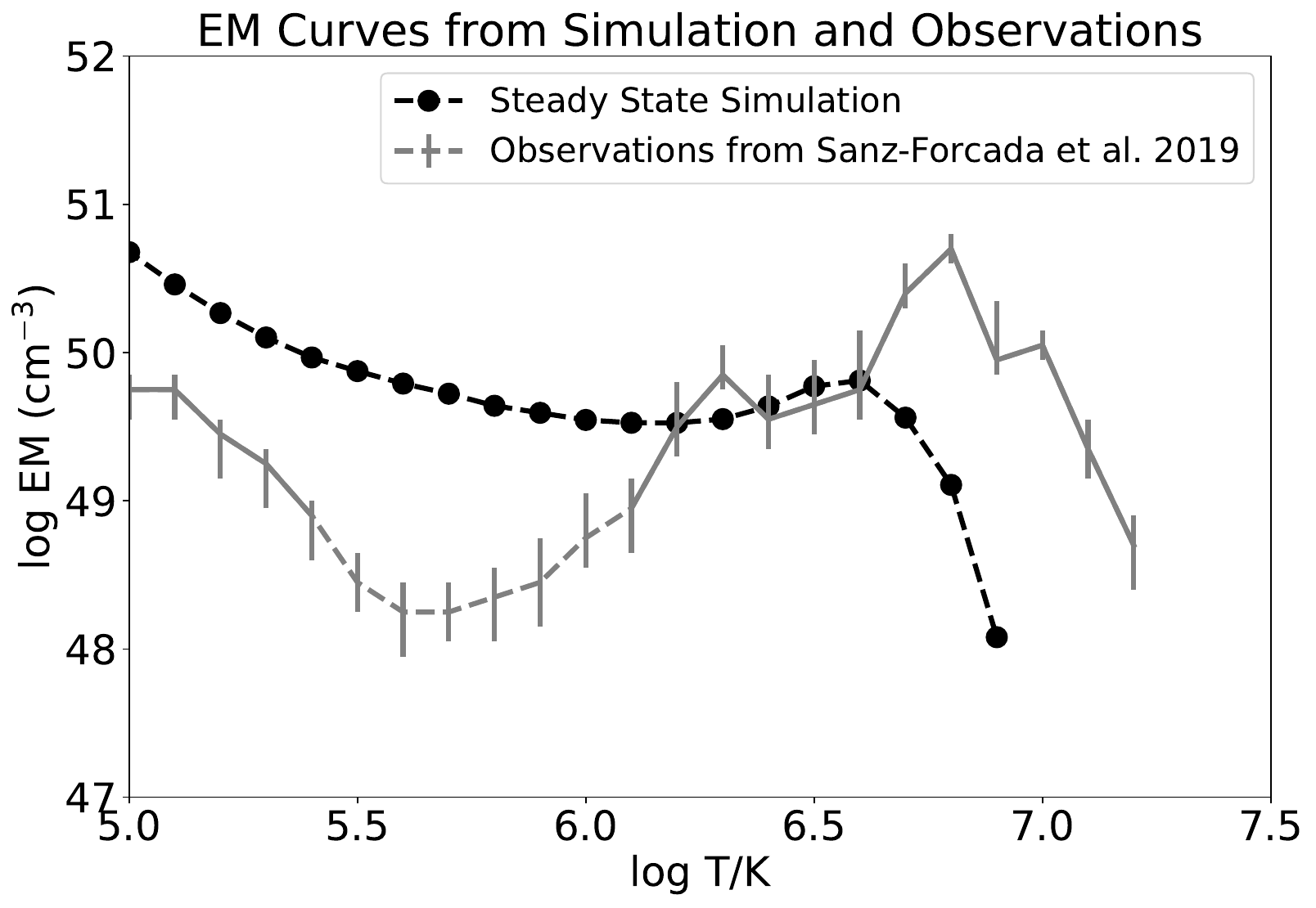}
    \caption{The EM distribution of the steady state model solution in this study (black) and from observations (grey). {The observed EM curve was adopted from \cite{SanzForcada2019} for the star $\iota$ Hor.}}
    \label{fig:emss}
\end{figure}

Figure \ref{fig:emss} plots the EM curve of our simulated stellar corona. {The EM curve obtained from spectroscopic observations in \cite{SanzForcada2019} for the star $\iota$ Hor was plotted as reference}. The simulated EM curve peaks at a temperature of around $10^{6.5}~\mathrm{K}$, indicating that our simulated corona has a temperature of around 3.2 MK. 

The simulated and observed EM distributions align well in the temperature range between $10^{6.1}~\mathrm{K}$ and $10^{6.6}~\mathrm{K}$. There is a conspicuous difference between the two curves in the higher temperature and lower temperature ranges. We interpret the lack of high-temperature plasma as possibly being due to (1) the limited spatial resolution in the simulation or insufficient heating in the model and/or (2) the extra high-temperature contribution in the observations being due to small flares below the detection threshold that are not included in the model. The overprediction of cooler plasma could also be related to the limited model resolution and an increase in thermal conduction due to poorly-resolved hot plasma structures. We emphasize, however, that reproducing the observed EM distribution is not a goal of this study and the exact EM distribution adopted is not important for the study in hand.

\subsection{CME events}
\begin{figure*}
    \centering
    \includegraphics[width=1.0\textwidth]{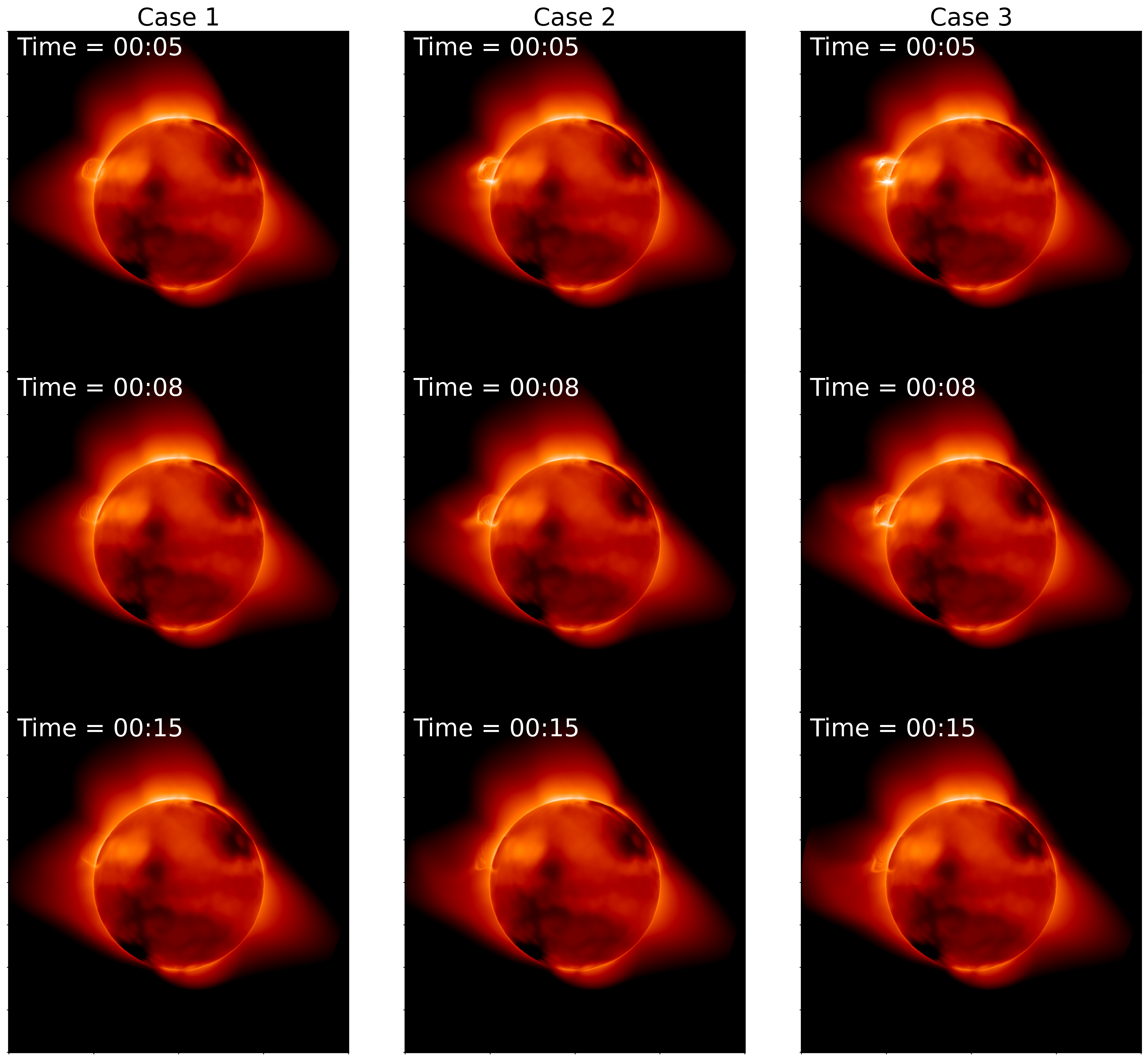}
    \caption{Snapshots of three cases at three time steps in XRT waveband. Each column shows the evolution from one case. The timestamp was labeled on each panel in the format of Hour:Minute. {The images are in the arbitrary unit and logarithmic scale. Movies are available.}.}
    \label{fig:cme3euv}
\end{figure*}

Figure \ref{fig:cme3euv} shows several snapshots of the three CME cases in the XRT waveband {, providing limb views of the three cases. We created animations in the eight bands from two viewing angles for each event. One viewing angle is the same as that in Fig. \ref{fig:cme3euv} while the other provides a disk view of the event. The animation follows the format of Fig. \ref{fig:euvss}, with the evolution time labeled on each frame as ''Time = Hour: Minute''.} Case 1 is the weakest while Case 3 has the most energetic CME among these three events. {The ejected material directing towards southern latitudes can be seen from the animations of Cases 2 and 3.} Case 1 is a weak event where no obvious mass escape was observed. 

We set criteria to capture and isolate the escaping masses in the 3D outputs of the model. The criteria are: (1) the speed of the mass should exceed the local escape speed, and (2) the radial velocity should be larger than the radial velocity in the steady state by at least 20 km/s. The first part requires the mass to have the ability to escape from the star. The second part captures the outward-moving mass after the initiation of the eruption while excluding the wind with a speed larger than the escape speed in the steady state. The "20 km/s" was determined by trial and error. The criteria have been used in our previous study \cite{Xu2024b} and proved the capability of isolating CME material.

In each case at each time step, the grids fulfilling the CME material criteria were selected and their total mass as well as average mass-weighted radial speed (i.e., radial bulk speed) were estimated. It is worth mentioning that in Case 1, although masses are moving away from the stellar surface, none of them reach their local escape speeds within the 45-minute simulation. Therefore, we referred to Case 1 as a confined event. The other two cases, Case 2 and 3, are eruptive. We averaged their masses, radial bulk velocities, and kinetic energies over the evolution time (i.e., 45 minutes) and the results are shown in Table. \ref{tab:cmes}. {Comparing these results with the masses and kinetic energies of solar CMEs \citep{Gopalswamy2009}, Case 2 falls within the typical solar range, while Case 3 is at the upper end of solar values. Here, we present only the averaged bulk velocities rather than the velocities at each time step because our CME selection is based on velocity characteristics, meaning the temporal evolution of these velocities does not carry significant physical meaning. Additionally, the kinematic properties of CMEs on young solar-type stars are beyond the scope of this work, so we did not further investigate their mass or velocity evolution. A previous study \citep{Xu2024b} examined the kinematic properties of ten simulated CMEs using similar simulations.}

\begin{table*}
    \centering
    \caption{CME Properties Averaged over the 45-Minute Simulation Time}
    \begin{tabular}{cccc}
    \hline
    \hline
        Case & Mass [g] & Radial Bulk Velocity [km/s] & Kinetic Energy [erg]\\
    \hline
        Case 2 & $4.26\times 10^{14}$ & 443 & $5.98\times 10^{29}$\\
        Case 3 & $1.14\times 10^{16}$ & 478 & $1.52\times 10^{31}$ \\
    \hline
    \end{tabular}
    \label{tab:cmes}
\end{table*}

\begin{figure}
    \centering
    \includegraphics[width=0.5\textwidth]{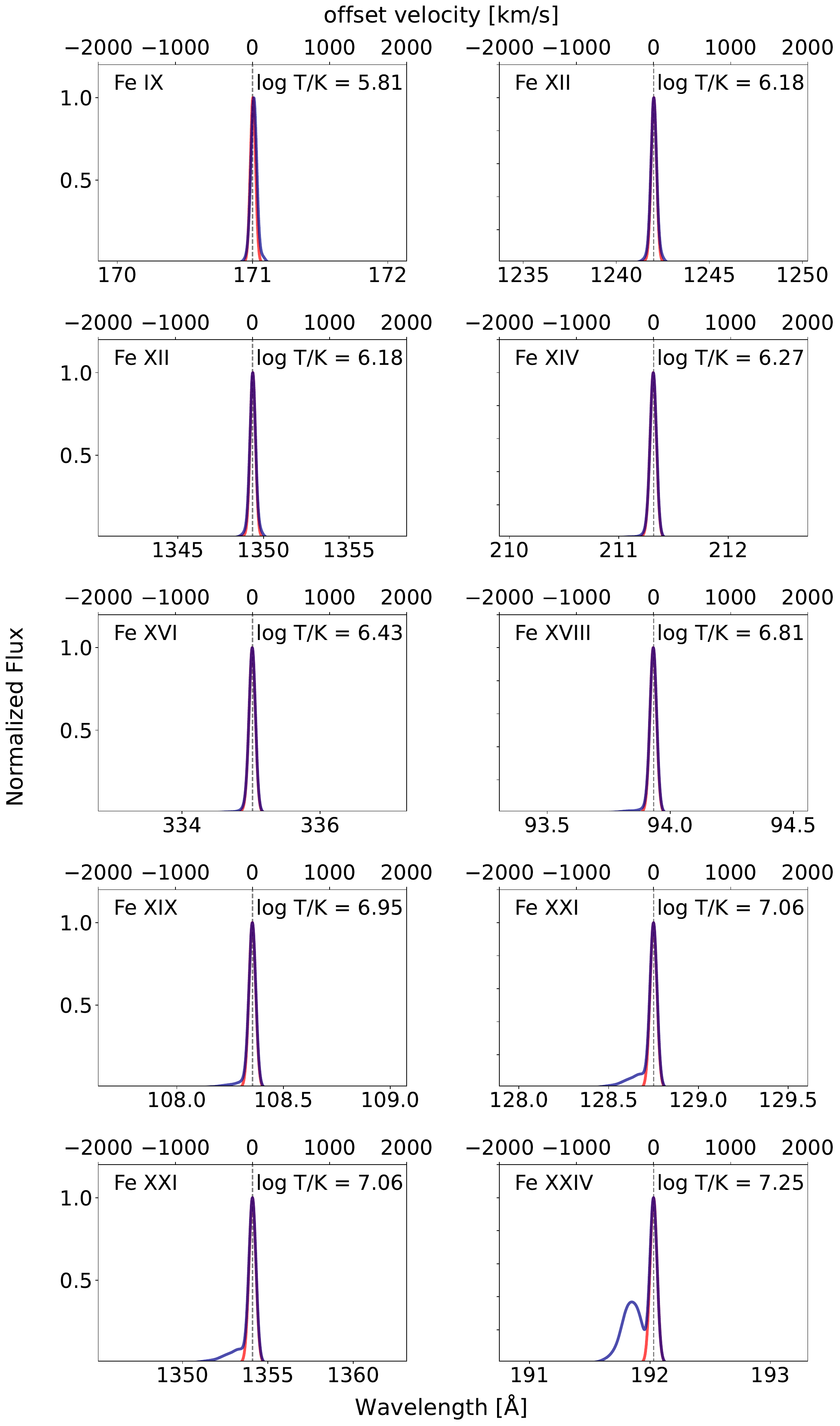}
    \caption{The synthetic line profiles at 8 minutes of evolution in Case 3 (blue solid curve). The red curves are the line profiles in the steady state. {In each panel, the blue and the red curves may partially overlap.} The vertical dashed grey lines mark the locations of the rest wavelengths of the lines. The ion name and line formation temperature are labeled on each panel. {All line profiles in the figure have infinite spectral resolution (i.e., no instrumental broadening).}}
    \label{fig:lines}
\end{figure}

Figure \ref{fig:lines} shows the line profiles (blue) in Case 3 at 8 minutes after the initiation of the eruption. {The observer was placed on the equatorial plane at the longitude where the flux rope was initiated. We selected this line of sight because the ejecta propagates (almost) along it, resulting in a larger projected velocity. This increases the likelihood of the CME component separating from the primary one, making it more detectable. This line of sight is perpendicular to that in Fig. \ref{fig:cme3euv}.} The corresponding ions and their formation temperatures are labeled on each panel. The red curves in the background are the line profiles in the steady state which have an infinite spectral resolution (i.e., $\sigma_{instru}=0$). Large asymmetries are apparent in the blue wings of the hotter lines. These lines have a temperature higher than around $10^{6.81}$ K (6.4 MK). The asymmetries last for 14 minutes after the initiation of the eruption in Case 3. Similar asymmetries were also observed in the line profiles in Case 1 and Case 2, but they lasted shorter than those in Case 3. The asymmetries lasted for 3 and 10 minutes in Case 1 and Case 2, respectively. It is worth noting that line Fe\,{\sc{xxi}} 128 \AA~and Fe\,{\sc{xxi}} 1354 \AA~have similar profiles because they are emitted by the same ion and have similar contribution functions. Therefore, in the following context, we only showed line profiles from Fe\,{\sc{xxi}} 1354 \AA~for simplicity. 

{We suggested that the observed asymmetries in the Case 2 and Case 3 are caused by the ejected material. The ejecta is a multi-temperature structure, which explains the different behaviors of different lines during the eruption. Seeing the animations of Case 2 and Case 3, the ejecta is more visible in high temperature channels (e.g., AIA 94 \AA, AIA 131 \AA, and XRT Al-mesh) than it in low temperature channels, indicating that most of the material has relatively high temperature. The slight red asymmetries in Fe\,{\sc{ix}} lines might be explained by the general downward motion of plasma behind the moving ejecta, which could be the result of the expansion of the ejecta. In the AIA 171 \AA~animation, there are loop-like structures with oscillations. They are cool material being trapped in the closed field line region. Their oscillations/fluctuations might be caused by the interaction of the ejected material and the backward plasma.}

We selected the five lines with clear asymmetries (i.e., Fe\,{\sc{xviii}} 94~\AA, Fe\,{\sc{xix}} 108~\AA, Fe\,{\sc{xxi}} 128~\AA/1354~\AA, and Fe\,{\sc{xxiv}} 192~\AA) for further analysis. We first added the instrumental broadening to the line profiles and sampled them based on the sampling principle introduced in Section \ref{sec:methods}. Then we integrated the profiles for each line over the first 20 minutes after the initiation of the eruption, aiming at mimicking an exposure time in real stellar observations. {The typical exposure time of the Extreme Ultraviolet Explorer (EUVE; \citealt{Bowyer1991}) ranged from around several thousands to several tens of thousand seconds with an effective area of around 1 $\mathrm{cm^{2}}$ \citep{Craig1997,Barstow2014}. The proposed future EUV spectrometer, such as the Extreme-ultraviolet Stellar Characterization for Atmospheric Physics and Evolution (ESCAPE; \citealt{France2022}), SIRIUS\citep{Barstow2012}, have effective areas of around one to two orders of magnitude larger than that of EUVE. Therefore, we chose an integration time of 20 minutes which is around 1--2 orders of magnitude shorter than the typical exposure time of EUVE.} 

\begin{figure*}
    \centering
    \includegraphics[width=1.0\textwidth]{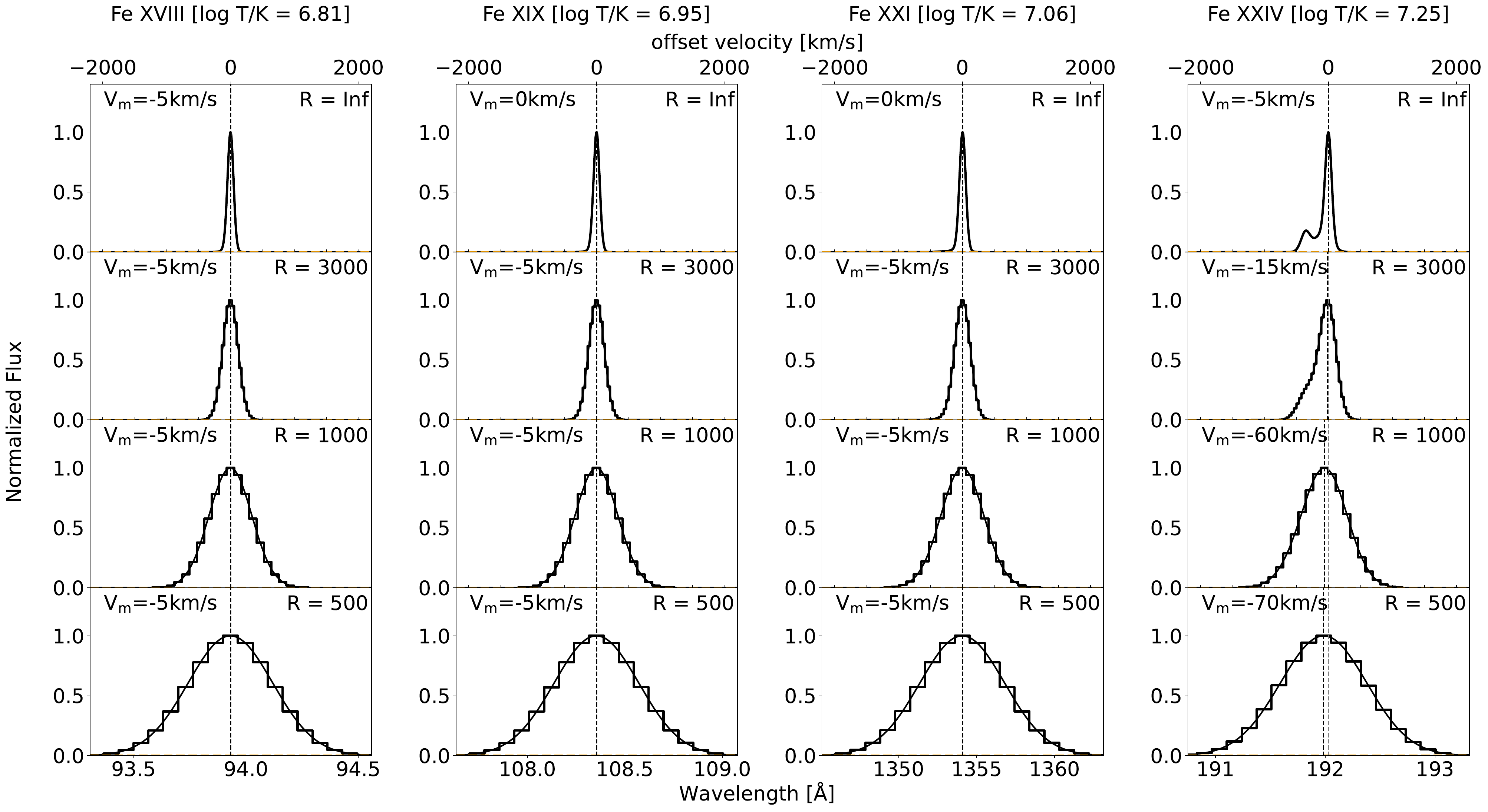}
    \caption{Synthetic line profiles for Case 1. The line profiles were integrated over the first 20 minutes of the evolution. The black curves are the synthetic line profiles and the steps are the results of sampling. The dashed grey vertical line indicates the rest wavelength of the line while the black vertical line represents {the location of the line centroid). The offset velocity of the line centroid is labeled as $v_m$ on each panel}. Panels in the same column show the line profiles of the same line but different spectral resolutions. The line information (i.e., ion name and line formation temperature) is labeled above each column. The spectral resolution, $R$, is labeled on each panel.}
    \label{fig:lines_wd1}
\end{figure*}

\begin{figure*}
    \centering
    \includegraphics[width=1.0\textwidth]{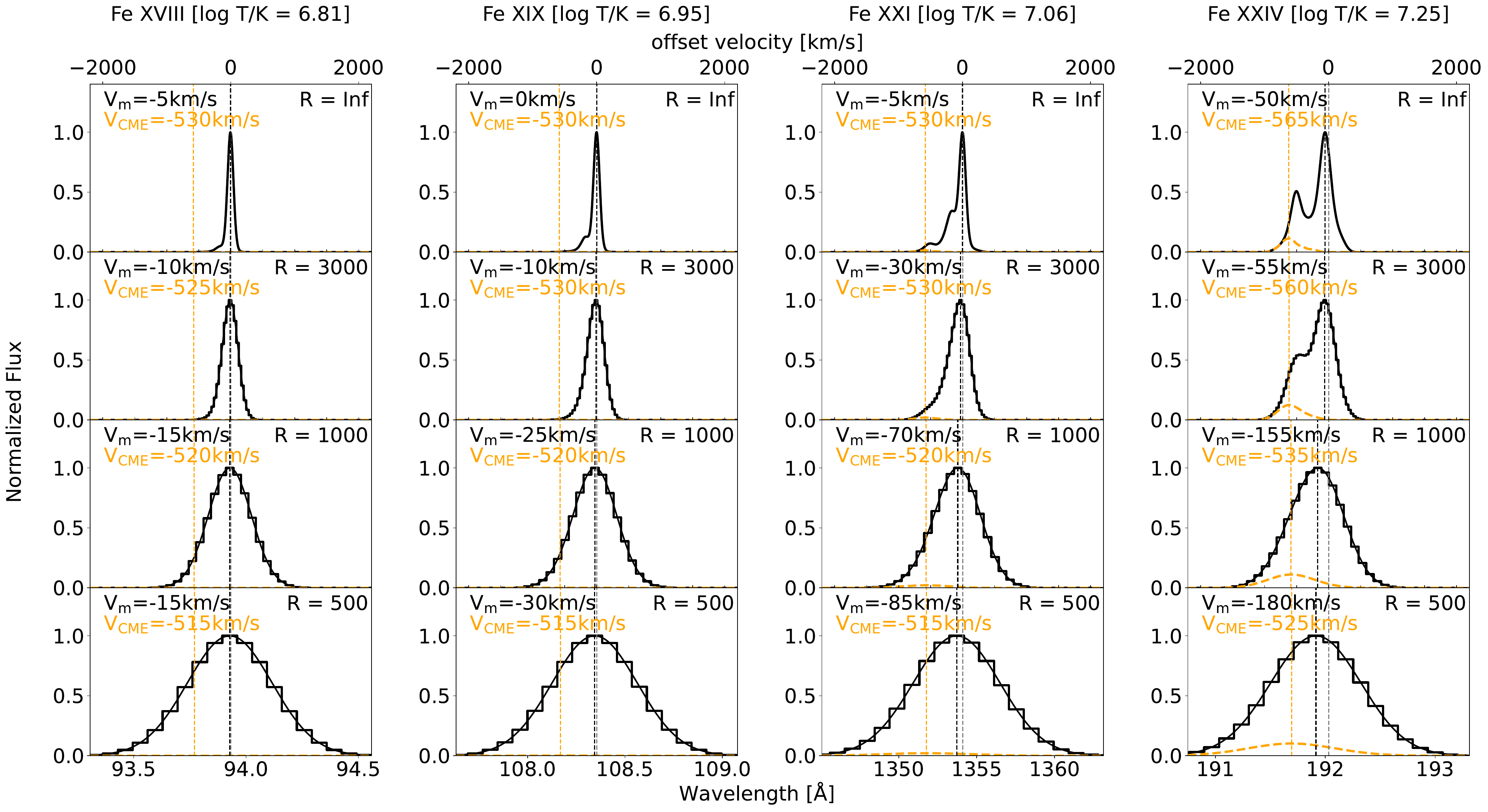}
    \caption{Similar to Fig. \ref{fig:lines_wd1} but for Case 2. The orange dashed curve shows the contribution from the escaping mass {which may be invisible if the contribution is weak. The location of the peak of the orange component is indicated by the vertical orange dashed line and the corresponding velocity is labeled as $\mathrm{v_{CME}}$}. }
    \label{fig:lines_wd2}
\end{figure*}

\begin{figure*}
    \centering
    \includegraphics[width=1.0\textwidth]{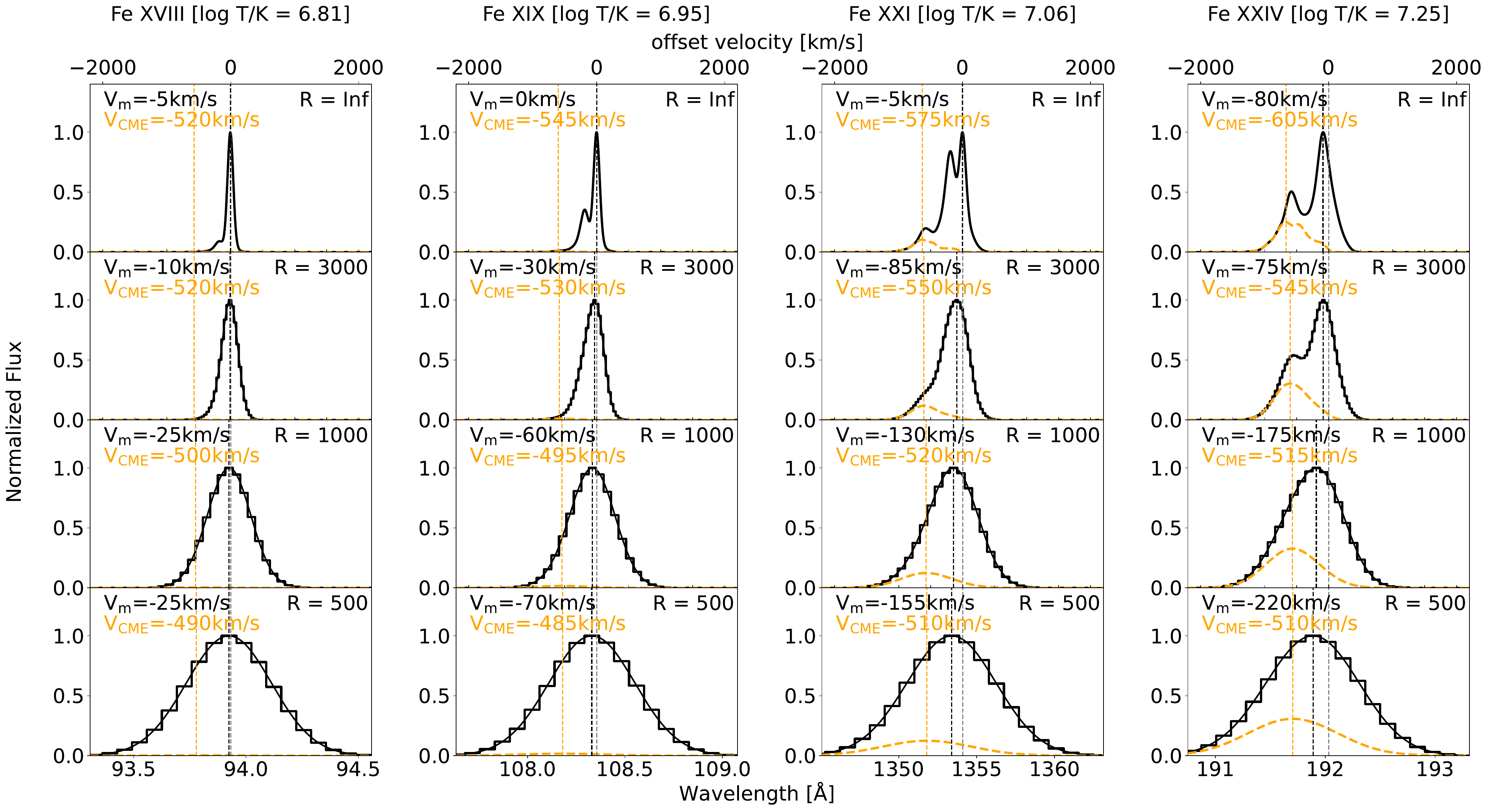}
    \caption{Similar to Fig. \ref{fig:lines_wd2} but for Case 3.}
    \label{fig:lines_wd3}
\end{figure*}

Figures \ref{fig:lines_wd1}--\ref{fig:lines_wd3} show the line profiles in each case under different spectral resolutions, $R$. We varied $R$ across the values [500,1000,2000,3000,4000,5000], but only present the results from $R=3000,1000,500$ for clarity. The black vertical lines in Figs. \ref{fig:lines_wd1}--\ref{fig:lines_wd3} mark the position of the {maximum of the line profile} (line centroid hereafter). The grey vertical lines indicate the rest wavelengths. As spectral resolution gets worse, the blue-shifted components are smoothed and become buried in the broadened primary components. In all three cases under the worst spectral resolution (i.e., $R = 500$), the asymmetries in the blue wings can barely be seen but the blue shifts of the line centroids are clearly visible in Fe\,{\sc{xxi}} and Fe\,{\sc{xxiv}} lines. {The offset velocities of the line centroids were labeled on each panel as $v_m$}. During the CME time, the blue shifts of the Fe\,{\sc{xxiv}} line centroids reach around -200 km/s in the two eruptive cases while in the confined case the offset velocity is around -70 km/s.

The orange components in Figs.~\ref{fig:lines_wd2} and \ref{fig:lines_wd3} show the contribution of the CME material captured using the criteria introduced at the beginning of this section. The orange vertical lines mark the wavelength of the maximum of the CME component and the corresponding offset velocities are labeled as $v_{CME}$ {which is by definition different from the radial bulk velocity listed in Tab. \ref{tab:cmes}}. We can see that the asymmetries in the blue wing are only partially contributed by CMEs. The rest is contributed by the mass that is accelerated during the eruption yet does not reach the local escape speed. This also applies to Case 1 where no CME is detected but there are asymmetries in the blue wings.

\begin{figure*}
    \centering
    \includegraphics[width=1.0\textwidth]{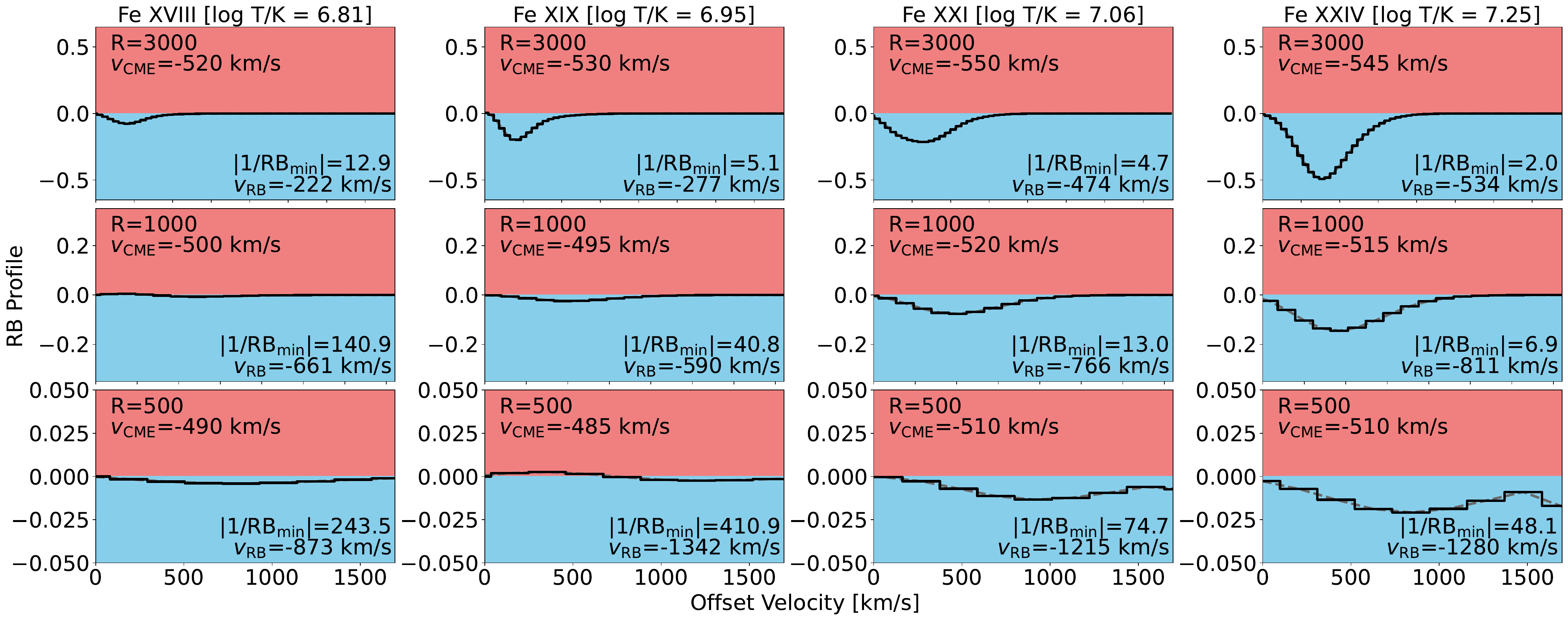}
    \caption{The RB profiles of Case 3 for four lines (i.e., Fe\,{\sc{xviii}} 94~\AA, Fe\,{\sc{xix}} 108~\AA, Fe\,{\sc{xxi}} 1354~\AA, and Fe\,{\sc{xxiv}} 192~\AA) under different spectral resolution ($R=3000, 1000, 500$). {The black steps are the consequence of sampling. The grey dotted curves represents the interpolation results of the black steps.} The ion name and its formation temperature are labeled over each column. The spectral resolution is also labeled in each panel. The $v_{\mathrm{CME}}$ in each panel is the velocity corresponding to the escaping mass and it is inherited from Fig. \ref{fig:lines_wd3}. The velocity $v_{\mathrm{RB}}$ corresponds to the local minimum of the RB profiles (i.e., $\mathrm{RB_{min}}$). {The minimum SNR required for detecting asymmetries is $|3\sqrt{2}/\mathrm{RB_{min}}|$.}}
    \label{fig:asymmetry_rb}
\end{figure*}

Figure \ref{fig:asymmetry_rb} presents RB profiles for Case 3. The RB asymmetry analysis was applied to all three cases on the integrated line profiles. The velocity $v_{RB}$ labeled on each panel corresponds to the local minimum of the RB profile (denoted as $RB_{min}$). As we can see, the asymmetry is more significant when the spectral resolution is higher and the asymmetries behave differently among different lines. We can now use the amplitude of the local minimum (i.e., $\mathrm{RB_{min}}$) to determine the minimum requirement for the noise level for the CME to be detected. We define the signal-to-noise ratio (SNR) as the peak intensity of the line (which is 1 in our studies) over the standard deviation of the noise (denoted as $\delta$). {Based on the calculation in Eq. \eqref{eq:rb}, the RB profile at each sampling point is obtained by subtracting the original signal, which results in the standard deviation of the noise in the RB profile being $\sqrt{2}\delta$.} We required the maximum amplitude of the asymmetries (i.e., $|\mathrm{RB_{min}}|$) to exceed three times the standard deviation of the noise in the RB profile, specifically $|\mathrm{RB_{min}}| > 3\sqrt{2}\delta$. Thus, the SNR (1/$\delta$) has a minimum requirement of $3\sqrt{2}/|\mathrm{RB_{min}}|$.

The minimum requirement of the SNR as a function of spectral resolution obtained using the RB profile approach is plotted in the upper row of Fig. \ref{fig:instrument}. {It is worth noting that we only considered the detectablity of the asymmetries. We did not take into account whether the velocities derived from the asymmetry techniques are comparable to the CME velocities.}

\begin{figure*}
    \centering
    \includegraphics[width=1.0\textwidth]{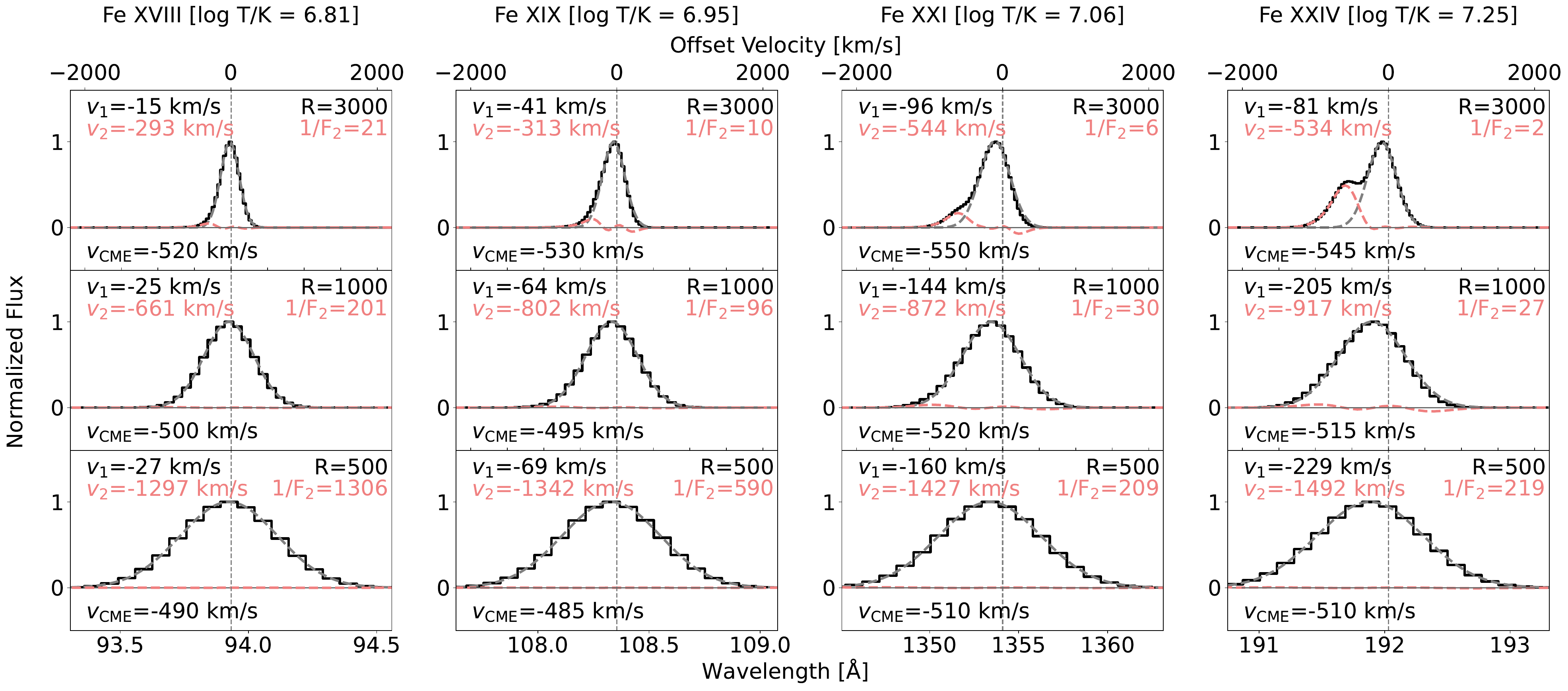}
    \caption{The single Gaussian fitting results of Case 3 for four lines (i.e., Fe\,{\sc{xviii}} 94~\AA, Fe\,{\sc{xix}} 108~\AA, Fe\,{\sc{xxi}} 1354~\AA, and Fe\,{\sc{xxiv}} 192~\AA) under different spectral resolution ($R=3000, 1000, 500$). The ion name and its formation temperature are labeled over each column and the spectral resolution is labeled in each panel. The $v_{\mathrm{CME}}$ in each panel is the velocity corresponding to the escaping mass and it is inherited from Fig.~\ref{fig:lines_wd3}. The velocity $v_1$ corresponds to the offset velocity of the fitted Gaussian, which is shown by the grey dashed curve. The velocity $v_2$ corresponds to the offset velocity of the peak of the residual which is shown by the red dashed curve. $\mathrm{F_2}$ is the peak value of the residual and {the minimum SNR required for detecting asymmetries is $|3/\mathrm{F_2}|$.} The grey vertical lines mark the rest wavelengths.}
    \label{fig:asymmetry_sgf}
\end{figure*}

\begin{figure*}
    \centering
    \includegraphics[width=0.9\textwidth]{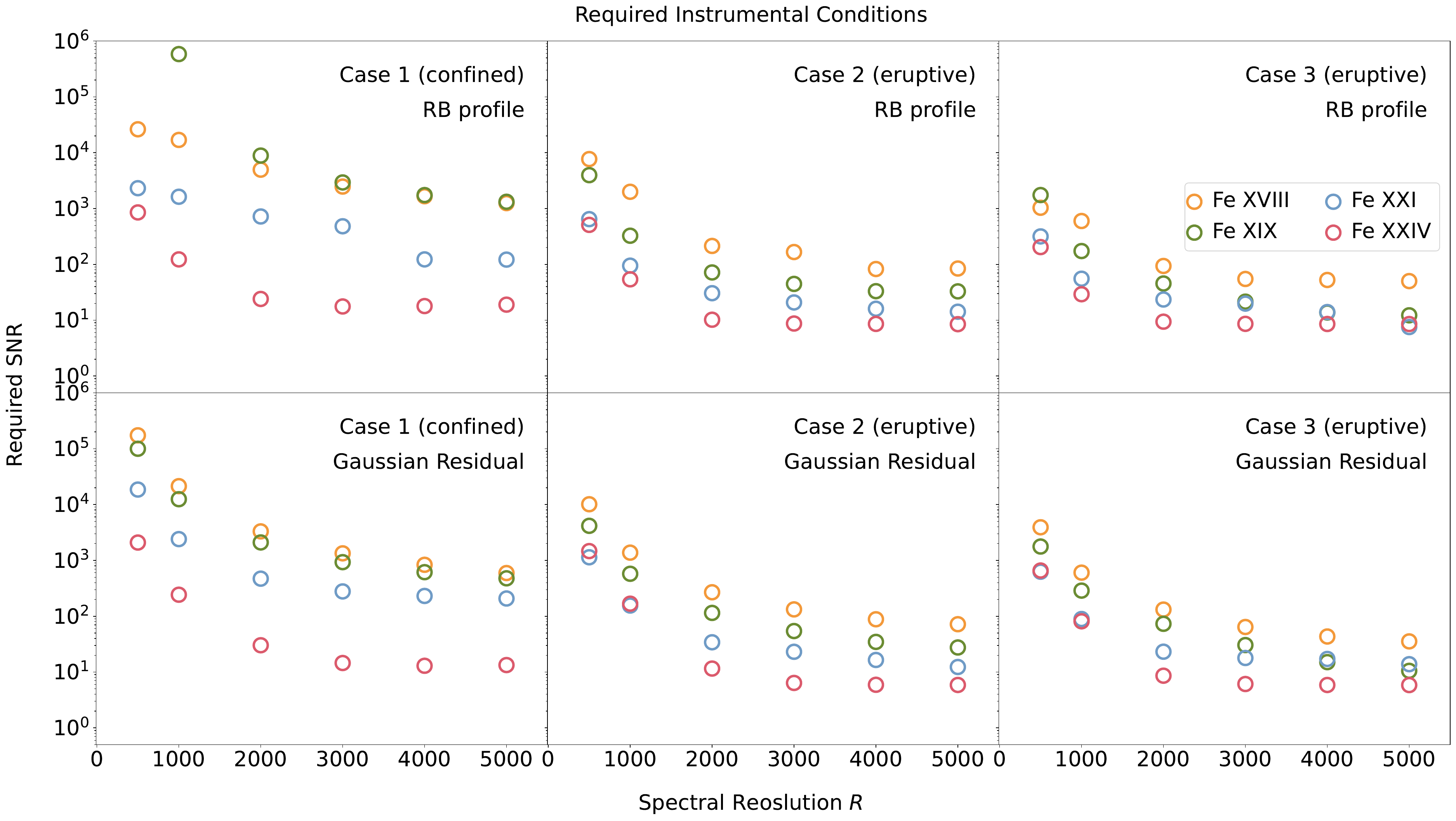}
    \caption{{The minimum requirements for the SNR under different spectral resolutions in three cases. Four colors mark the results using four ions. Panels in the upper row are results from the RB profile technique while those in the bottom row are from the Gaussian residual method.}}
    \label{fig:instrument}
\end{figure*}

Figure \ref{fig:asymmetry_sgf} shows the results from the single Gaussian residual method for Case 3. The velocity $\mathrm{v_2}$ corresponds to the local maximum of the residual (denoted as $\mathrm{F_2}$). {Assuming that the Gaussian fitting process does not introduce any additional noise}, similar to the results in Fig.~\ref{fig:asymmetry_rb}, the minimum requirement of the SNR is $3/\mathrm{F_2}$ and the results are plotted in the bottom row of Fig.~\ref{fig:instrument}. 

{The asymmetries arise from material disturbed during the eruption. The background wind also contributes during the eruption, affecting both the primary component and the residual. If the wind is accelerated but not to a very high speed, its effect may be embedded within the primary component, making it subject to fitting and subtraction. This explains why the velocity of the primary component, $v_1$, is not zero during the eruption. As spectral resolution decreases, $v_1$ tends to increase, and the primary component broadens to account for more of the accelerated wind's contribution. If the wind undergoes significant acceleration but does not reach the local escape velocity (i.e., it does not meet the CME criteria), its contribution separates from the primary component and appears in the residual. As a result, the residual includes both the CME (as defined by the CME criteria) and the significantly accelerated wind. The velocity, $v_2$, represents a combination of these two contributions. For this reason, we do not compare $v_2$ directly with $v_{\text{CME}}$. However, it's important to note that asymmetries at the extreme blue end should be attributed solely to CMEs, as their velocity properties satisfy our CME criteria. Additionally, under appropriate spectral resolution, some lines exhibit similar $v_2$ and $v_{\text{CME}}$ values (e.g., $R = 3000$ in Fe {\sc{xxi}} and Fe {\sc{xxiv}}). This occurs because the primary component is broad enough to encompass the accelerated wind, while the spectral resolution remains high enough to preserve asymmetries. A similar explanation applies to the discrepancy between the velocity derived from RB analysis (i.e., $v_{\text{RB}}$) and $v_{\text{CME}}$. The RB profile contains contributions from both the accelerated wind and the CME material. Therefore, $v_{\text{RB}}$ is not necessarily equal to $v_{\text{CME}}$.}

Figure \ref{fig:instrument} plots the minimum requirement of SNR against the spectral resolution in three cases using the two aforementioned methods. The different colors show the results from different lines. There is a general trend that the larger asymmetries require less stringent instrumental performance. The larger asymmetries exist in: (1) higher spectral resolution, (2) eruptive events where materials reach escape velocities compared to the confined event, and (3) lines with formation temperatures close to the dominant temperature of the escaping material. It is also worth noting that if the resolution is above 2000, the increase of the spectral resolution does not further reduce the requirement for the SNR.

\section{Discussion}\label{sec:discussion}
{Small blueshifts during flares and mass ejection events have been found in some observations of Extreme ultraviolet Variability Experiment (EVE; \citealt{Woods2012}) onboard SDO (e.g., in He {\sc{ii}} 30.4 nm in \citealt{Chamberlin2016}; in H Lyman series in \citealt{Brown2016}; Cheng, Z., et al., submitted to ApJ), although the formation temperatures of the EVE lines are 1-2 orders of magnitude lower than ours. The stellar X-ray observations of CME candidates in \cite{Argiroffi2019} and \cite{Chen2022} showed blue-shifted line profiles without significant asymmetries in their blue wings. \cite{Argiroffi2019} discovered significant blueshifts in the lines S {\sc{xvi}} (4.73 \AA, $T\sim25$ MK) and Si {\sc{xiv}} (6.18 \AA, $T\sim16$ MK) during the impulsive phase of the flare, with velocities of around 300 km/s, as well as blueshifts in the line O {\sc{viii}} (18.97 \AA, $T\sim3$ MK) after the flare, with velocities of around 100 km/s. The former were interpreted as hot plasma moving within the flare loop, while the latter were suggested to be caused by cool CME material moving upward. {\cite{Chen2022} detected Doppler shifts in the lines O {\sc{viii}} (18.97 \AA, $T\sim3$ MK), Fe {\sc{xvii}} (15.01 \AA, $T\sim6$ MK), Mg {\sc{xii}} (8.42 \AA, $T\sim10$ MK), and Si {\sc{xiv}} (6.18 \AA, $T\sim16$ MK)}, with velocities of up to 130 km/s during several flares on EV Lac, which were interpreted as chromospheric evaporation. During one flare, cool ($\sim$3 MK) and warm (5–10 MK) upflows, accompanied by a decrease in plasma density, were suggested to be indicative of a stellar filament/prominence eruption. {These observations, obtained by Chandra X-ray Observatory space telescope \citep{Weisskopf2000,Canizares2005}, have a spectral resolution of around 1000.} The behavior of lines in these observations is similar to our synthetic line profiles with spectral resolutions of $R=1000$ and $R=500$, where the blue-shifted components are not resolved. {The velocities of the blue-shifted line centroids are around -100 to -200 km/s in the two eruptive events in our simulations.} They are smaller than the velocities of the CME components, as shown in Fig. \ref{fig:lines_wd2} and \ref{fig:lines_wd3} by the orange profiles, whose velocities reach around -500 km/s. This suggests that when the spectral resolution is insufficient, the blue shifts of the line centroids might serve as an indicator of stellar CMEs, and the velocities of the shifted centroids are the lower limits of the LOS velocities of the CMEs.}

{Apart from the CME material, the plasma accelerated but with velocities smaller than the escape velocities also contributes to the asymmetries in the line profiles. This is to say, there are intrinsic discrepancies between the CME LOS velocities and the velocities derived from the asymmetries. Thus, we did not make comparisons between them. However, if the CME component contributes to most of the asymmetries and the spectral resolution is moderate, the velocities derived from the asymmetries may be comparable to the CME velocities. For instance, in line profiles of Fe {\sc{xxiv}} in Case 3, when the spectral resolution is 3000, the CME velocity is -545 km/s and the velocities obtained from the RB analysis and the single Gaussian residual technique are both -534 km/s.}

The total broadening of lines contributed by the thermal and non-thermal parts is on the order of several tens km/s, which suggests that increasing the spectral resolution wouldn't help to make the line more narrow after the spectral resolution reaches around 3000. This can be supported by investigating the requirements for the SNR using spectra without instrumental broadening. The minimum requirements of SNR without instrumental broadening are similar to those with spectral resolutions higher than 3000.

{The change of the ratio between the primary and the secondary components will affect the estimation of the critical SNR or the spectral resolution. Such changes happen if we took other factors into consideration. For example, the flare is not well captured in the model, which may affect the intensities of the primary component in the line profiles. Also, as we mentioned before, there is a possible underestimation of the high temperature material in the stellar corona, which is reflected by the difference between the simulated and the observed EM curves in the high temperature regime. This may result in much stronger background components (i.e., the primary components) in the high temperature lines, causing the critical SNR to be larger. Other factors, such as the viewing angles and the integration time, also affect the presence of the line profiles and thus, the critical SNR and the spectral resolution.} However, it is important to note that the absolute values of SNR or spectral resolution from this study may not be directly applicable to instrument design. Our goal is to demonstrate that the blue wing asymmetries caused by CMEs could be visible in XUV lines under certain conditions (i.e., combination of suitable CME properties and instrumental conditions). We also suggested a method for using MHD simulation results to synthesis line profiles of optically thin lines, analyze the asymmetries in line profiles, and obtain constraints for instrumentation. However, for these results to be useful in future instrument designs, more advanced/refined MHD simulations and a broader exploration of CME properties are necessary. {In reality, multiple CMEs could occur simultaneously and their interaction will also change the line profiles. However, this effect is beyond the scope of this work and will be investigated in the future.}

{Last but not least, we focused on deriving velocities from the synthetic spectra but paid little attention to CME mass, as we believe the current simulation results do not allow for a reliable mass estimation. First, determining mass from EUV lines (or other optically thin lines) is challenging. }

{As far as we know, there are two main methods for deriving CME mass from full-disk EUV spectroscopic observations. The first involves using the relationship between coronal dimming and CME mass (e.g., \citealt{Mason2016, Loyd2022}) based on solar observations and/or theoretical derivations. However, these relationships remain debated, especially as the dataset expands. Also, in our simulations, different lines exhibit varying dimming amplitudes, observed only in lines with formation temperatures higher than Fe {\sc{xiv}}. However, determining the dimming amplitude is challenging, as it requires a baseline. Initially, we used steady-state intensities, but the flux rope insertion altered line intensities by adding plasma across a wide temperature range, affecting nearly all lines except Fe {\sc{xxiv}}. Using post-insertion intensities as a baseline is also problematic, as the system is not yet in equilibrium, leading to greater deviation of EM from observations. While dimming is present in some lines, its amplitude remains uncertain. Moreover, CMEs are multi-thermal structures, meaning dimming in a few lines does not fully represent their total mass. Therefore, we did not adopt this method.}

{The second method of deriving mass, presented in \cite{Argiroffi2019}, assumes a radiative cooling time to determine the electron density of the ejecta. This estimated electron density is then used to derive the CME volume from the observed EM, allowing for an estimation of the CME mass. In their case, the assumed radiative cooling time (of approximately 200 ks) was based on the duration over which asymmetries were observed in the O {\sc{viii}} line. However, our simulation covers only 45 minutes, and the system has not yet returned to a steady state. Therefore, we did not adopt this method. Deriving CME mass from EUV observations is an interesting and complex topic that requires more detailed investigation, which we plan to explore in future work.}

\section{Summary}\label{sec:summary}
We simulated CMEs on a young solar-type star whose parameters were inspired by the well-studied $\iota$ Horologii. We created three eruption cases with different initial energies and allowed them to evolve for 45 minutes. One of them is a confined case and the other two are eruptive events. The masses, radial bulk velocities, and kinetic energies for the two eruptive cases are summarized in Table \ref{tab:cmes}.

We synthesized line profiles of several EUV lines based on the 3D outputs of the simulations. {Asymmetries were observed in the blue wings of lines during the eruptions.} We investigated the asymmetries in the blue wings of lines during the eruptions under different spectral resolutions. The asymmetries are more significant in (1) eruptive cases, (2) higher spectral resolution, and (3) lines with formation temperature close to the temperature of the ejecta material. {In the situation with a spectral resolution higher than around 2000, the secondary peaks related to the plasma motion are clearly visible in some lines, such as Fe {\sc{xxiv}}. Line profiles with a spectral resolution lower than around 2000 do not show significant asymmetries, but their line centroids shift to blue wings, which are considered one of the indicators of plasma motion.} Two methods were used to quantify the amplitudes of the asymmetries, from which we obtained some requirements for instrumentation.

{Our study shows that the blue wing asymmetries or blue shifts of line centroids caused by CMEs can be detected in disk-integrated XUV line profiles under certain conditions. We proposed that more refined MHD simulations and detailed analysis are necessary in order for results of such studies to be applicable in future instrumentation.}

\section*{acknowledgement}
This work is supported by the National Natural Science Foundation of China grants 12425301, 12250006, and 12073004. This work is carried out using the SWMF/BATSRUS tools developed at the University of Michigan Center for Space Environment Modeling (CSEM) and made available through the NASA Community Coordinated Modeling Center (CCMC). H. T. is grateful for the support from the New Cornerstone Science Foundation through the Xplorer Prize. 

\newpage
\bibliographystyle{aasjournal}
\bibliography{ref}

\end{document}